\documentclass[aps,superscriptaddress,10pt,prl,twocolumn]{revtex4}
\usepackage{amsmath}
\usepackage{graphicx}
\usepackage{color}
\usepackage{natbib}
\usepackage{pifont}
\usepackage{hyperref}
\usepackage{subfigure}

\newcommand{\LR}[1]{\left(#1\right)}

\begin{document}

\title{Multimode circuit QED with hybrid metamaterial transmission lines}
\author{D. J. Egger}
\affiliation{Theoretical Physics, Universit\"at des Saarlandes, Saarbr\"ucken, Germany}
\author{F. K. Wilhelm}
\affiliation{Theoretical Physics, Universit\"at des Saarlandes, Saarbr\"ucken, Germany}
\affiliation{IQC and Department of Physics and Astronomy, University of Waterloo, ON, Canada}

\begin{abstract}
Quantum transmission lines are central to superconducting and hybrid quantum
computing. In this work we show how coupling them to a left-handed transmission line
 allows circuit QED to reach a new regime: multi-mode
ultra-strong coupling. Out of the many
potential applications of this novel device, we discuss the preparation of
multipartite entangled states and the simulation of
the spin-boson model where a quantum phase
transition is reached up to finite size effects. 
\end{abstract}

\maketitle

Quantum optics addresses the interaction of quanta of matter ---
atoms --- with quanta of electromagnetic fields --- photons. This is
beautifully realized in cavity quantum electrodynamics (QED) \cite{Haroche06}, where the
interaction between those units is made strong by confining the field
into a small mode volume \cite{Blais04}. Circuit QED takes this further by
confining microwave photons in a quasi 1D strip-line cavity and using
superconducting qubits as {\em artificial} atoms with a large dipole
moment \cite{Blais04,Schoelkopf08}. Next to being a promising architecture for quantum computing,
a multitude of basic quantum optical effects has been
demonstrated \cite{You11}. Going beyond what can be reached in atomic
systems, an ultrastrong coupling regime --- where the coupling strength
becomes comparable to the atomic energy scales --- has been proposed
 \cite{Bourassa09} and
achieved \cite{Forndiaz10,Niemczyk10}.
Furthermore in the circuit QED approach, elements are
entirely human-made and can hence be flexibly engineered. This can lead to 
coupling to multiple modes \cite{Filipp11,Switch,NOON,Wang11,Mariantoni11b}
either in the same or distinct cavities. There is a wealth of
proposals exploiting these features to create complex photonic states
 \cite{Nunnenkamp11,Hartmann06,Underwood12} involving a large number of cavities. 
Parallel to these
developments are those of left-handed meta-materials. 
They have a wide variety of applications in
photonics from the microwave to the visible range such as
invisibility cloaks and perfect flat lenses \cite{Veselago68,Pendry00}. For classical guided microwaves,
left-handed transmission lines have been proposed \cite{Eleftheriades02} and studied \cite{Salehi05} on the macroscopic scale.
In the following we show how a hybrid transmission line, made of left and right-handed media, coupled to a flux qubit gives rise to
ultrastrong multimode coupling.

\begin{figure*}[htbp!]
 \includegraphics[width=0.8\textwidth]{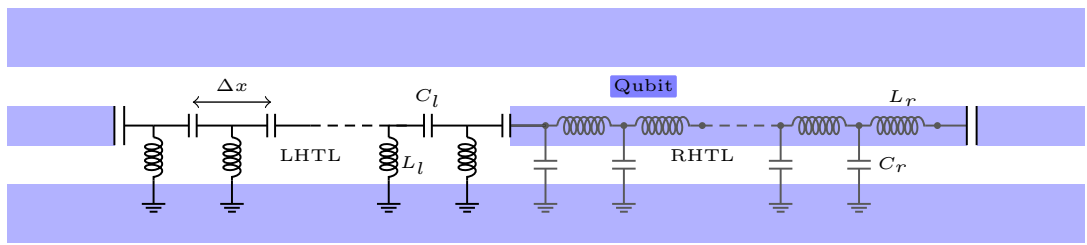}
 \caption{(color online). Discrete LHTL coupled to a continuous RHTL \label{Fig:LumpedCTL}. The regular right-handed part of the
   transmission line is on the right, connected to a left-handed line
   shown on the left. The terminating capacitors allow to externally access the
   modes. The qubit couples to the right-handed component. In the supplementary material it is shown that stray capacitance to ground in
 the left handed line does not change the physics. Additional stray inductances and capacitances can be taken into account as shown by Eleftheriades, {\em et al.} \cite{Eleftheriades02}. The light blue areas indicate strip-lines and ground planes.}
\end{figure*}

{\em The System:} In one-dimension, left-handedness is defined as the wave vector $\mathbf{k}_l$ and the
Poynting vector having opposite orientation; the phase and
group-velocity are opposite corresponding to a falling dispersion relation $\partial\omega(k)/\partial k<0$.
This can be achieved \cite{Eleftheriades02} by a discrete array of series capacitors and
parallel inductors to ground, see Fig.  \ref{Fig:LumpedCTL}. A low loss left-handed transmission line (LHTL)
can be realized with superconductors \cite{Salehi05,Jung_arXiv}. This is
the dual (inductors and capacitors interchanged) of the usual \cite{Pozar05}
discrete representation of the right-handed
transmission line (RHTL). In
practice, the LHTL remains a metamaterial composed of discrete elements, whereas
the RHTL is a metal strip 
represented as the continuum limit of a ladder network \cite{Pozar05}.
We can understand the physics of this line as follows:
For any ladder network with discrete time-translation symmetry, the
eigenmodes are (propagating or decaying) plane waves with a dispersion
relation derived from the solutions of \cite{Pozar05} 
\begin{equation}
\sin \LR{\frac{k\Delta x}{2}} =\pm\frac12 i\sqrt{\frac{Z_s}{Z_p}}
\label{eq:planewave}
\end{equation}
For a RHTL, substituting  impedances of
the series elements $Z_s=i\omega L_r$ and parallel elements
$Z_p=(i\omega C_r)^{-1}$ gives the usual dispersion
\begin{equation}
 \omega_r\left(k_r\right)=~\frac{2}{\sqrt{C_rL_r}}\sin\LR{\frac{k_r\Delta
     x}{2}} \xrightarrow[\text{continuum}]{}\frac{k_r}{\sqrt{c_rl_r}} 
\end{equation}
For the
LHTL, we interchange the roles of inductors and capacitors and
obtain from equation (\ref{eq:planewave}) propagating modes (real-valued
$k$ for real $\omega$) with the opposite dispersion
relation 
\begin{equation}
 \omega_l\left(k_l\right)=~\frac{1}{2\sqrt{C_lL_l}\sin\LR{\frac{k_l\Delta x}{2}}} \label{Eqn:Dispertion_L}
\end{equation}
Here, $C_{l/r}$ and $L_{l/r}$ are capacitances and inductances as defined in
Fig. \ref{Fig:LumpedCTL} and $c_r$ and $l_r$ are the capacitance and
inductance per unit length in the RHTL. $\Delta x$ is the size of a unit cell.
More details are in the supplementary material. 

\begin{figure}
     \includegraphics[width=0.98\columnwidth]{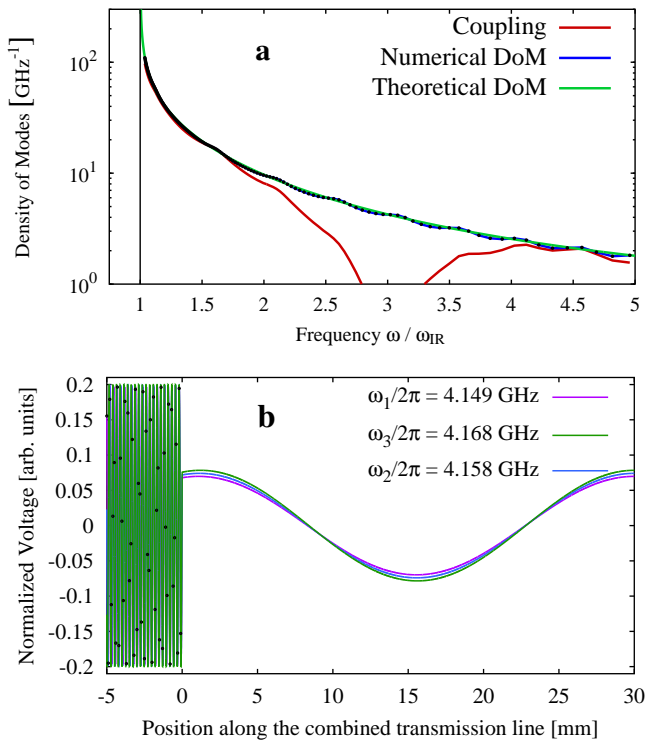}
     \caption{(color online). (a) Example of density of modes, showing a lower band-edge at
     $\omega_{\rm IR}$. Dots indicate actual modes, the green curve the
     approximate formula which is in excellent agreement. The red curve shows the coupling strength between a $0.5$ mm long flux qubit placed at a current anti-node of the transmission line, as in Fig. \ref{Fig:LumpedCTL}. Designing the qubit to couple to the $4.579~\mathrm{GHz}$ mode with strength $460~\mathrm{MHz}$, results in ultrastrong-coupling to 50 modes within a $\pm 460~\mathrm{MHz}$ range. (b) Example of the first three consecutive normal modes, the voltage profiles in the RHTL are almost identical. The LHTL has 200 unit cells measuring $100~\mathrm{\mu m}$ each. In the LHTL the voltages at the discrete unit cells, for the first mode only, are shown by the black dots; the continuous lines serve as a guide to the eye to see the mode structure. Requiring a $50~\Omega$ impedance and an IR cutoff at $\omega_\text{IR}/2\pi=4$ GHz sets $C_l=398~\mathrm{fF}$ and $L_l=995~\mathrm{pH}$.
     The parameters for the RHTL were chosen so that it supports a full wavelength at $\omega_\text{IR}$. This sets the values for its total inductance and capacitance. Therefore a $3~\mathrm{cm}$ long RHTL requires a capacitance  and inductance per unit length of $c_r=1667~\mathrm{fF/\mu m}$ and $l_r=4167~\mathrm{pH/\mu m}$. \label{Fig:DoSExample}}
 \end{figure}

Unusual physics arises when right- and left-handed media are
interfaced \cite{Veselago68}. We realize this with a coupled transmission line (CTL) shown in
Fig. \ref{Fig:LumpedCTL}, a discrete LHTL coupled to a RHTL to be
taken into the continuum limit. 
A key unusual feature of the LHTL, as compared to a
regular RHTL, is the divergence of the density of modes
(DoM) at a low-frequency bound $\omega_\text{IR}=1/2\sqrt{C_lL_l}$,
seen in Fig. \ref{Fig:DoSExample}\textcolor{blue}{(a)}, implying the
existence of a quasi-continuous band even in a cavity.
In the LHTL, low frequencies correspond to short wavelengths
due to the falling dispersion relation $\omega(k)$. Thus, by only a
small change in frequency, a new orthogonal mode can be found that is
different by one node in the left-handed component.
As the wavelength approaches the lattice
constant, the dispersion relation in equation (\ref{Eqn:Dispertion_L})
becomes flat due to Bragg reflection \cite{Ashcroft76} --- the 
DoM develops a van-Hove-type singularity setting the aforementioned
divergence at $\omega_{\rm IR}$.
Due to the hybrid nature of this new CTL, the closely spaced
frequencies at this lower band-edge have nearly-identical spatial
structures in the RHTL. The fast oscillation in the LHTL ensure orthogonality between modes.
Figure \ref{Fig:DoSExample}\textcolor{blue}{(b)} shows three consecutive low frequency
modes obtained from the full solution. Close to $\omega_\text{IR}$ the RHTL provides a mere
constant contribution thus the DoM is dominated by the divergence due
to the LHTL and vice versa. In consequence, 
the DoM can be approximated by the sum of the densities in the
uncoupled lines
\begin{equation} \notag
 \mathcal{D}\LR{\omega}
 =\frac{4N_l\sqrt{C_lL_l}}{\pi}\tan\phi_L\sin\phi_L,\quad
 \phi_L=\frac{k_l(\omega)\Delta x}{2}.
\label{eq:approx_DOM}
\end{equation}
$N_l$ is the number of cells in the LHTL. The agreement between this prediction and the numerically obtained
modes of the full model is excellent up to small oscillations, see
Figure \ref{Fig:DoSExample}. To engineer the DOM, one can control
$\omega_{\rm IR}$ by the mesh size and independently $N_l$ by the
length of the LHTL.

To provide good coupling between both components one would like to
have an impedance $Z_0$ (typically $50~\Omega$) requiring $C_l=(2\omega_\text{IR}Z_0)^{-1}$ and
$L_l=Z_0/2\omega_\text{IR}$. Furthermore for coupling to qubits
$\omega_\text{IR}$ should be chosen to lie around qubit frequencies (e.g. $4$ GHz).
The capacitances could be realized with interdigitated as
well as with overlap capacitors and the parallel inductors could be
realized with Josephson Junctions in the linear regime since they
provide sufficient inductance in a small footprint. Furthermore it is shown in the supplementary material that disorder in these parameters has little effect.

The quantum behavior of the CTL is obtained
through canonical quantization of the circuit in Fig. \ref{Fig:LumpedCTL}.
This leads to a system of uncoupled quantum harmonic oscillators, each
described by operators $\hat a^\dagger_n$,$\hat a_n^{\phantom{\dagger}}$ acting on modes with frequencies $\omega_n$.
A qubit described in its energy eigenbasis by Pauli matrices $\hat{\sigma}_{x/z}$ 
placed close to the CTL will couple to mode $n$ with strength $g_n$
\begin{equation}
\hat{H}/\hbar=\frac{\Delta_0}{2}\hat\sigma_z+\sum_n
g_n\hat\sigma_x\left(\hat{a}_n^{\phantom{\dagger}}+\hat{a}_n^\dagger\right)
+\sum_n \omega_n \hat{a}^\dagger_n\hat{a}_n^{\phantom{\dagger}}\,.
\label{eq:multiham}
\end{equation}
If the qubit is coupled to the RHTL, $g_n\simeq g_{n+1}$ for low 
frequency modes since they have similar spatial profiles in the RHTL.
For a flux qubit \cite{Mooij99,Insight}, the mode dependent part of the coupling strength is given
by $\mathcal{D}(\omega_n)\langle I_n(x)
\rangle/\mathrm{max}_n\{\langle I_n(x) \rangle\}$. 
The current $I_n $ is averaged over the spatial extent of the qubit. Figure \ref{Fig:DoSExample}
shows that the qubit can be coupled to a wide range of modes. For
frequencies sufficiently above $\omega_{IR}$ the wavelength in the
RHTL also starts to change away from the antinode towards a node, creating a deep minimum in coupling strength.
This mode structure allows the qubit to simultaneously couple to multiple modes when $N>1$ modes fall within a frequency
interval of $2g_n$. We refer to this regime as {\em multi-mode
strong-coupling}. It can be reached with other superconducting qubits, notably
transmons \cite{Koch07}, which should be placed at a charge
antinode. Flux qubits on the other hand allow us to reach {\em multi-mode ultrastrong
coupling} \cite{Bourassa09,Forndiaz10} --- $g_n/\omega_n>0.1$. This regime offers many new possibilities for circuit QED.

\begin{figure}[htbp!] \centering
  \includegraphics[width=0.48\textwidth]{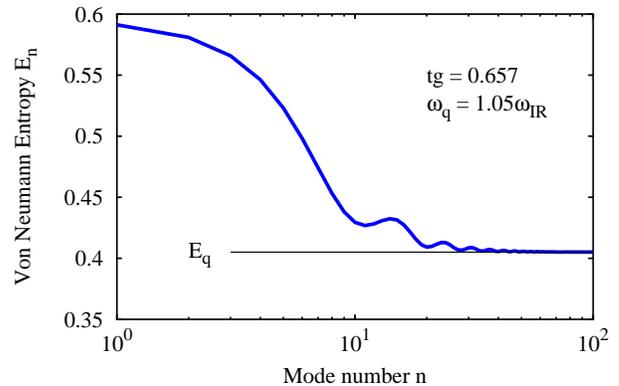}
  \caption{(color online). Von Neumann entropy as function of the traced out mode
    $n$. Here the qubit is put in the bath at a frequency
    $\omega_q>\omega_\text{IR}$ and the system left to evolve for a
    dimensionless time $tg$. The Von Neumann entropy is computed for
    the system after having traced out the qubit and mode $n$. We then
    vary $n$ to see how much the entropy increases from $E_q$, the
    entropy when only the qubit is traced out. $E_q>0$ indicates that
    their is at least bi-partite entanglement. When additionally
    tracing out mode $n$ the entropy $E_n$ increase above $E_q$ and
    this for all modes. The latter indicates complex multi-partite
    entanglement.\label{Fig:Entropy2}} 
\end{figure}

{\em Applications:} The multimode Rabi Hamiltonian, equation
(\ref{eq:multiham}), allows us to prepare multimode
entangled states. Within the rotating wave approximation it 
conserves the number of excitations.
Exciting the qubit and placing its $0\leftrightarrow1$
resonance frequency slightly above $\omega_{\rm IR}$ allows the qubit
excitation to distribute itself over many modes, i.e.,  produce arbitrary
superpositions of the form $c_0|1;0\rangle+\sum_n c_n |0;n\rangle$.
$|0;n\rangle$ indicates the qubit in
the ground state, a single photon in mode $n$ and none in the other
modes. For $|1;0\rangle$ only the qubit is excited.
These states are in general entangled as seen from their Von Neumann entropy \cite{Guhne09}.
Figure \ref{Fig:Entropy2} shows the entropy, indicating multimode entanglement, of
a single excitation, starting in $|1;0\rangle$, that spread out over all modes
for a time $t$.

The Spin-Boson model \cite{Leggett87} is a fundamental model of
quantum dissipation which allows to understand the transition between
coherent and incoherent behaviour as well as a quantum phase transition
suppressing quantum tunneling. It is described by the Hamiltonian in equation 
(\ref{eq:multiham}) in the limit where the modes form a
continuum. The dense modes at the low-frequency end provide a
generic and 
realizable quantum simulator for this model.
Our unusual density of modes provides
a novel regime of sub-subhomic models with a low-frequency cutoff, i.e., a
spectral density of the form
\begin{equation} \notag
J(\omega)=\sum_ng_n^2\delta(\omega-\omega_n)\simeq\frac{N_l}{\pi\sqrt{2\omega_\text{IR}}}\frac{\Theta(\omega-\omega_\text{IR})}{\sqrt{\omega-\omega_\text{IR}}}.
\end{equation}
The ground and lowest excited states are well approximated by a multimode
Schr\"odinger cat state of the qubit dressed by coherent photonic
states \cite{Leggett87}
\begin{equation}
|\pm\rangle=\frac{1}{\sqrt{2}}\left(|L\rangle\bigotimes_n
  |\lambda_n\rangle \pm|R\rangle\bigotimes_n |-\lambda_n\rangle
\right).
\label{eq:dressedstate}
\end{equation}
$|L,R\rangle$ are the eigenstates of $\hat\sigma_x$. The renormalized energy splitting is
\begin{equation}
\Delta_{\rm eff}=\left\langle +\left|\hat{H}\right|
  -\right\rangle=\Delta_0\exp\bigg(-2\sum_n\lambda_n^2\bigg).
\label{eq:scalegap}
\end{equation}
The multimode cat state in equation (\ref{eq:dressedstate}) involves,
according to the principle of adiabatic renormalization, all fast modes
, those with $\omega_n>\Delta_{\rm eff}$, as they can adiabatically follow the qubit.
Slow modes remain unaffected. Thus $\lambda_n=\frac{g_n^2}{\omega_n^2}\Theta(\omega_n-\Delta_{\rm
  eff})$ which leads to a self-consistency relation for $\Delta_{\rm
  eff}$. The ratio $\Delta_{\rm eff}/\Delta_0$ 
measures the accumulated phase space distance of the dressing clouds,
i.e. the total cat size \cite{Haroche06}, by taking the logarithm of equation
(\ref{eq:scalegap}). Thus, the low-energy states of the system are
strongly renormalized as are their effective energies.

\begin{figure}[htbp!] \centering
 \includegraphics[width=0.48\textwidth]{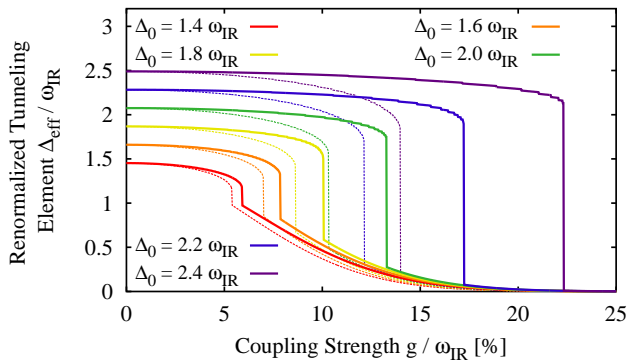}
 \caption{(color online). Plot of the renormalized tunnelling as function of the global coupling strength $g$. 
 The individual mode coupling $g_n$ is the product of $g$ and the mode-dependent spatial contribution bounded by unity. The thin dashed lines correspond to the case when this effect is neglected i.e. $\langle I_n(x)\rangle/\max_n\{\langle I_n(x)\rangle\}=1$.
 This allows simultaneous tuning of the coupling to all modes. The discontinuous drop in $\Delta_\text{eff}$ distinguishes the regions of weakly and strongly renormalized energy splitting. \label{Fig:RenormTunel}}
\end{figure}

A true dissipative quantum phase transition \cite{Leggett87,Weiss99}
has $\Delta_{\rm eff}=0$ in the localised phase. This limit would be reached if the modes were infinitely
close (hence arbitrarily close to $\omega_{\rm IR}$) as would result from an infinitely long LHTL or if $\omega_{\rm IR}\rightarrow 0$
as in the case of infinitely dense LHTL unit cells. Note
that in the usual sub-Ohmic spin-boson model, the latter is
assumed. We thus conclude that our system approaches a quantum phase
transition in the infinite sample limit.

\begin{figure}[htbp!] \centering
 \includegraphics[width=0.48\textwidth]{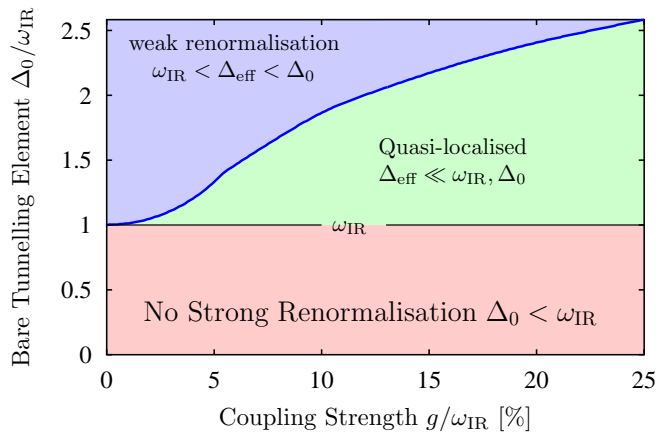}
 \caption{(color online). Finite-size phase diagram of the qubit-CTL system. The
   density of states for the case at hand is the same is in
   Fig. \ref{Fig:DoSExample} with the dip in coupling taken into account. \label{Fig:Renorm} }
\end{figure}

To corroborate the finite size-behaviour, we have studied the ground and first excited state of the qubit-CTL
model using its actual modes in the adiabatic renormalization
approach.  We identify
multiple regimes: for weak coupling or large $\Delta_0/\omega_{\rm IR}$, there is
only weak dressing manifest by a small shift of $\Delta_0$. At stronger
coupling, we observe the quasi-localized phase, with $\Delta_{\rm
  eff}\ll \Delta_0$. 
Remarkably, even at finite length, the two
regimes are separated by a discontinuous transition as indicated by
Figure \ref{Fig:RenormTunel}. 
Figure \ref{Fig:Renorm} shows the corresponding finite-size phase diagram
highlighting the need for ultrastrong coupling.
We see that by tuning the bare qubit frequency slightly above the cutoff,
$\Delta_0>\omega_{\rm IR}$ \cite{Paauw09} we can tune the system through the phase
transition {\em in situ}, or by employing a tunable coupler. The phase transition is manifest by a
discontinuous drop in the energy splitting (as measured through
spectroscopy) of the qubit that is inconsistent with the tuning of the
circuit alone, see FIg. \ref{Fig:RenormTunel}. Engineering the
transmission line to have dense enough modes and an appropriate
$\omega_{IR}$ can be accomplished using equation (\ref{eq:approx_DOM}) and
$\omega_{IR}=1/2\sqrt{C_lL_l}$. 

On the level of partition functions, this model is equivalent to a one-dimensional Ising chain
\cite{Cardy81,Weiss99}. This is discussed in more detail in the
supplementary material.
In the present case, this would be an Ising chain with an
interaction that decays as $\propto |i-j|^{-1/2}$, where $i$ and $j$
are site indices, up to a range $r\propto \omega_{\rm IR}^{-1}$, after which it decays
exponentially. Thus, when cooled from high-temperatures the system
is well described by mean-field theory, which predicts a
ferromagnetic phase transition, until the correlation length reaches
$r$. At that point, the system follows short-range physics and remains
paramagnetic between magnetized blocks of size $r$ --- in analogy to the tunnel coupling in
the spin-boson model falling deeply, but not to zero.

In conclusion, we have proposed an engineered hybrid transmission line
that allows to reach a new multimode strong coupling regime of circuit
QED by combining a regular line with a
metamaterial. This will open the way for novel applications in
microwave photonics and strongly correlated photon states, out of
which we have outlined the generation of multimode entanglement, 
multimode Schr\"odinger cat states and quantum phase transitions. 

We acknowledge useful discussions with Heiko Rieger and Britton
Plourde. We thank Emily Pritchett for her careful reading of the
manuscript. This work was funded in parts by DARPA through the QuEST
program and by the European Union through ScaleQIT.

\bibliographystyle{apsrev}
\bibliography{Lines_Bib}{}

\begin{thebibliography}{30}
\expandafter\ifx\csname natexlab\endcsname\relax\def\natexlab#1{#1}\fi
\expandafter\ifx\csname bibnamefont\endcsname\relax
  \def\bibnamefont#1{#1}\fi
\expandafter\ifx\csname bibfnamefont\endcsname\relax
  \def\bibfnamefont#1{#1}\fi
\expandafter\ifx\csname citenamefont\endcsname\relax
  \def\citenamefont#1{#1}\fi
\expandafter\ifx\csname url\endcsname\relax
  \def\url#1{\texttt{#1}}\fi
\expandafter\ifx\csname urlprefix\endcsname\relax\def\urlprefix{URL }\fi
\providecommand{\bibinfo}[2]{#2}
\providecommand{\eprint}[2][]{\url{#2}}

\bibitem[{\citenamefont{Haroche and Raimond}(2006)}]{Haroche06}
\bibinfo{author}{\bibfnamefont{S.}~\bibnamefont{Haroche}} \bibnamefont{and}
  \bibinfo{author}{\bibfnamefont{J.-M.} \bibnamefont{Raimond}},
  \emph{\bibinfo{title}{Exploring the Quantum: Atoms, Cavities, and Photons}}
  (\bibinfo{publisher}{Oxford University Press}, \bibinfo{address}{Oxford},
  \bibinfo{year}{2006}).

\bibitem[{\citenamefont{Blais et~al.}(2004)\citenamefont{Blais, Huang,
  Wallraff, Girvin, and Schoelkopf}}]{Blais04}
\bibinfo{author}{\bibfnamefont{A.}~\bibnamefont{Blais}},
  \bibinfo{author}{\bibfnamefont{R.-S.} \bibnamefont{Huang}},
  \bibinfo{author}{\bibfnamefont{A.}~\bibnamefont{Wallraff}},
  \bibinfo{author}{\bibfnamefont{S.~M.} \bibnamefont{Girvin}},
  \bibnamefont{and} \bibinfo{author}{\bibfnamefont{R.~J.}
  \bibnamefont{Schoelkopf}}, \bibinfo{journal}{Phys. Rev. A}
  \textbf{\bibinfo{volume}{69}}, \bibinfo{pages}{062320}
  (\bibinfo{year}{2004}).

\bibitem[{\citenamefont{Schoelkopf and Girvin}(2008)}]{Schoelkopf08}
\bibinfo{author}{\bibfnamefont{R.}~\bibnamefont{Schoelkopf}} \bibnamefont{and}
  \bibinfo{author}{\bibfnamefont{S.}~\bibnamefont{Girvin}},
  \bibinfo{journal}{Nature} \textbf{\bibinfo{volume}{451}},
  \bibinfo{pages}{664} (\bibinfo{year}{2008}).

\bibitem[{\citenamefont{You and Nori}(2011)}]{You11}
\bibinfo{author}{\bibfnamefont{J.}~\bibnamefont{You}} \bibnamefont{and}
  \bibinfo{author}{\bibfnamefont{F.}~\bibnamefont{Nori}},
  \bibinfo{journal}{Nature} \textbf{\bibinfo{volume}{474}},
  \bibinfo{pages}{589} (\bibinfo{year}{2011}).

\bibitem[{\citenamefont{Bourassa et~al.}(2009)\citenamefont{Bourassa, Gambetta,
  Abdumalikov, Astafiev, Nakamura, and Blais}}]{Bourassa09}
\bibinfo{author}{\bibfnamefont{J.}~\bibnamefont{Bourassa}},
  \bibinfo{author}{\bibfnamefont{J.~M.} \bibnamefont{Gambetta}},
  \bibinfo{author}{\bibfnamefont{A.~A.} \bibnamefont{Abdumalikov}},
  \bibinfo{author}{\bibfnamefont{O.}~\bibnamefont{Astafiev}},
  \bibinfo{author}{\bibfnamefont{Y.}~\bibnamefont{Nakamura}}, \bibnamefont{and}
  \bibinfo{author}{\bibfnamefont{A.}~\bibnamefont{Blais}},
  \bibinfo{journal}{Phys. Rev. A} \textbf{\bibinfo{volume}{80}},
  \bibinfo{pages}{032109} (\bibinfo{year}{2009}).

\bibitem[{\citenamefont{Forn-Diaz et~al.}(2010)\citenamefont{Forn-Diaz,
  Lisenfeld, Marcos, Garcia-Ripoll, Solano, Harmans, and Mooij}}]{Forndiaz10}
\bibinfo{author}{\bibfnamefont{P.}~\bibnamefont{Forn-Diaz}},
  \bibinfo{author}{\bibfnamefont{J.}~\bibnamefont{Lisenfeld}},
  \bibinfo{author}{\bibfnamefont{D.}~\bibnamefont{Marcos}},
  \bibinfo{author}{\bibfnamefont{J.~J.} \bibnamefont{Garcia-Ripoll}},
  \bibinfo{author}{\bibfnamefont{E.}~\bibnamefont{Solano}},
  \bibinfo{author}{\bibfnamefont{C.~J. P.~M.} \bibnamefont{Harmans}},
  \bibnamefont{and} \bibinfo{author}{\bibfnamefont{J.~E.} \bibnamefont{Mooij}},
  \bibinfo{journal}{Phys. Rev. Lett.} \textbf{\bibinfo{volume}{105}},
  \bibinfo{pages}{237001} (\bibinfo{year}{2010}).

\bibitem[{\citenamefont{Niemczyk et~al.}(2010)\citenamefont{Niemczyk, Deppe,
  Huebl, Menzel, Hocke, Schwarz, Garcia-Ripoll, Zueco, H\"ummer, Solano
  et~al.}}]{Niemczyk10}
\bibinfo{author}{\bibfnamefont{T.}~\bibnamefont{Niemczyk}},
  \bibinfo{author}{\bibfnamefont{F.}~\bibnamefont{Deppe}},
  \bibinfo{author}{\bibfnamefont{H.}~\bibnamefont{Huebl}},
  \bibinfo{author}{\bibfnamefont{E.~P.} \bibnamefont{Menzel}},
  \bibinfo{author}{\bibfnamefont{F.}~\bibnamefont{Hocke}},
  \bibinfo{author}{\bibfnamefont{M.~J.} \bibnamefont{Schwarz}},
  \bibinfo{author}{\bibfnamefont{J.~J.} \bibnamefont{Garcia-Ripoll}},
  \bibinfo{author}{\bibfnamefont{D.}~\bibnamefont{Zueco}},
  \bibinfo{author}{\bibfnamefont{T.}~\bibnamefont{H\"ummer}},
  \bibinfo{author}{\bibfnamefont{E.}~\bibnamefont{Solano}},
  \bibnamefont{et~al.}, \bibinfo{journal}{Nat. Phys.}
  \textbf{\bibinfo{volume}{6}}, \bibinfo{pages}{772} (\bibinfo{year}{2010}).

\bibitem[{\citenamefont{Filipp et~al.}(2011)\citenamefont{Filipp, G\"oppl,
  Fink, Baur, Bianchetti, Steffen, and Wallraff}}]{Filipp11}
\bibinfo{author}{\bibfnamefont{S.}~\bibnamefont{Filipp}},
  \bibinfo{author}{\bibfnamefont{M.}~\bibnamefont{G\"oppl}},
  \bibinfo{author}{\bibfnamefont{J.~M.} \bibnamefont{Fink}},
  \bibinfo{author}{\bibfnamefont{M.}~\bibnamefont{Baur}},
  \bibinfo{author}{\bibfnamefont{R.}~\bibnamefont{Bianchetti}},
  \bibinfo{author}{\bibfnamefont{L.}~\bibnamefont{Steffen}}, \bibnamefont{and}
  \bibinfo{author}{\bibfnamefont{A.}~\bibnamefont{Wallraff}},
  \bibinfo{journal}{Phys. Rev. A} \textbf{\bibinfo{volume}{83}},
  \bibinfo{pages}{063827} (\bibinfo{year}{2011}).

\bibitem[{\citenamefont{Mariantoni et~al.}(2008)\citenamefont{Mariantoni,
  Deppe, Marx, Gross, Wilhelm, and Solano}}]{Switch}
\bibinfo{author}{\bibfnamefont{M.}~\bibnamefont{Mariantoni}},
  \bibinfo{author}{\bibfnamefont{F.}~\bibnamefont{Deppe}},
  \bibinfo{author}{\bibfnamefont{A.}~\bibnamefont{Marx}},
  \bibinfo{author}{\bibfnamefont{R.}~\bibnamefont{Gross}},
  \bibinfo{author}{\bibfnamefont{F.~K.} \bibnamefont{Wilhelm}},
  \bibnamefont{and} \bibinfo{author}{\bibfnamefont{E.}~\bibnamefont{Solano}},
  \bibinfo{journal}{Phys. Rev. B} \textbf{\bibinfo{volume}{78}},
  \bibinfo{pages}{104508} (\bibinfo{year}{2008}).

\bibitem[{\citenamefont{Merkel and Wilhelm}(2010)}]{NOON}
\bibinfo{author}{\bibfnamefont{S.~T.} \bibnamefont{Merkel}} \bibnamefont{and}
  \bibinfo{author}{\bibfnamefont{F.~K.} \bibnamefont{Wilhelm}},
  \bibinfo{journal}{New J. Phys.} \textbf{\bibinfo{volume}{12}},
  \bibinfo{pages}{093036} (\bibinfo{year}{2010}).

\bibitem[{\citenamefont{Wang et~al.}(2011)\citenamefont{Wang, Mariantoni,
  Bialczak, Lenander, Lucero, Neeley, O'Connell, Sank, Weides, Wenner
  et~al.}}]{Wang11}
\bibinfo{author}{\bibfnamefont{H.}~\bibnamefont{Wang}},
  \bibinfo{author}{\bibfnamefont{M.}~\bibnamefont{Mariantoni}},
  \bibinfo{author}{\bibfnamefont{R.~C.} \bibnamefont{Bialczak}},
  \bibinfo{author}{\bibfnamefont{M.}~\bibnamefont{Lenander}},
  \bibinfo{author}{\bibfnamefont{E.}~\bibnamefont{Lucero}},
  \bibinfo{author}{\bibfnamefont{M.}~\bibnamefont{Neeley}},
  \bibinfo{author}{\bibfnamefont{A.~D.} \bibnamefont{O'Connell}},
  \bibinfo{author}{\bibfnamefont{D.}~\bibnamefont{Sank}},
  \bibinfo{author}{\bibfnamefont{M.}~\bibnamefont{Weides}},
  \bibinfo{author}{\bibfnamefont{J.}~\bibnamefont{Wenner}},
  \bibnamefont{et~al.}, \bibinfo{journal}{Phys. Rev. Lett.}
  \textbf{\bibinfo{volume}{106}}, \bibinfo{pages}{060401}
  (\bibinfo{year}{2011}).

\bibitem[{\citenamefont{Mariantoni et~al.}(2011)\citenamefont{Mariantoni, Wang,
  Bialczak, Lenander, Lucero, Neeley, O'Connell, Sank, Weides, Wenner
  et~al.}}]{Mariantoni11b}
\bibinfo{author}{\bibfnamefont{M.}~\bibnamefont{Mariantoni}},
  \bibinfo{author}{\bibfnamefont{H.}~\bibnamefont{Wang}},
  \bibinfo{author}{\bibfnamefont{R.~C.} \bibnamefont{Bialczak}},
  \bibinfo{author}{\bibfnamefont{M.}~\bibnamefont{Lenander}},
  \bibinfo{author}{\bibfnamefont{E.}~\bibnamefont{Lucero}},
  \bibinfo{author}{\bibfnamefont{M.}~\bibnamefont{Neeley}},
  \bibinfo{author}{\bibfnamefont{A.~D.} \bibnamefont{O'Connell}},
  \bibinfo{author}{\bibfnamefont{D.}~\bibnamefont{Sank}},
  \bibinfo{author}{\bibfnamefont{M.}~\bibnamefont{Weides}},
  \bibinfo{author}{\bibfnamefont{J.}~\bibnamefont{Wenner}},
  \bibnamefont{et~al.}, \bibinfo{journal}{Nat. Phys.}
  \textbf{\bibinfo{volume}{7}}, \bibinfo{pages}{287} (\bibinfo{year}{2011}).

\bibitem[{\citenamefont{Nunnenkamp et~al.}(2011)\citenamefont{Nunnenkamp, Koch,
  and Girvin}}]{Nunnenkamp11}
\bibinfo{author}{\bibfnamefont{A.}~\bibnamefont{Nunnenkamp}},
  \bibinfo{author}{\bibfnamefont{J.}~\bibnamefont{Koch}}, \bibnamefont{and}
  \bibinfo{author}{\bibfnamefont{S.~M.} \bibnamefont{Girvin}},
  \bibinfo{journal}{New J. Phys.} \textbf{\bibinfo{volume}{13}},
  \bibinfo{pages}{095008} (\bibinfo{year}{2011}).

\bibitem[{\citenamefont{Hartmann et~al.}(2006)\citenamefont{Hartmann, Brandao,
  and Plenio}}]{Hartmann06}
\bibinfo{author}{\bibfnamefont{M.}~\bibnamefont{Hartmann}},
  \bibinfo{author}{\bibfnamefont{F.}~\bibnamefont{Brandao}}, \bibnamefont{and}
  \bibinfo{author}{\bibfnamefont{M.}~\bibnamefont{Plenio}},
  \bibinfo{journal}{Nat. Phys.} \textbf{\bibinfo{volume}{2}},
  \bibinfo{pages}{849} (\bibinfo{year}{2006}).

\bibitem[{\citenamefont{Underwood et~al.}(2012)\citenamefont{Underwood, Shanks,
  Koch, and Houck}}]{Underwood12}
\bibinfo{author}{\bibfnamefont{D.~L.} \bibnamefont{Underwood}},
  \bibinfo{author}{\bibfnamefont{W.~E.} \bibnamefont{Shanks}},
  \bibinfo{author}{\bibfnamefont{J.}~\bibnamefont{Koch}}, \bibnamefont{and}
  \bibinfo{author}{\bibfnamefont{A.~A.} \bibnamefont{Houck}},
  \bibinfo{journal}{Phys. Rev. A} \textbf{\bibinfo{volume}{86}},
  \bibinfo{pages}{023837} (\bibinfo{year}{2012}).

\bibitem[{\citenamefont{Veselago}(1968)}]{Veselago68}
\bibinfo{author}{\bibfnamefont{V.}~\bibnamefont{Veselago}},
  \bibinfo{journal}{Sov. Phys. Usp.} \textbf{\bibinfo{volume}{10}},
  \bibinfo{pages}{517} (\bibinfo{year}{1968}).

\bibitem[{\citenamefont{Pendry}(2000)}]{Pendry00}
\bibinfo{author}{\bibfnamefont{J.~B.} \bibnamefont{Pendry}},
  \bibinfo{journal}{Phys. Rev. Lett.} \textbf{\bibinfo{volume}{85}},
  \bibinfo{pages}{3966} (\bibinfo{year}{2000}).

\bibitem[{\citenamefont{Eleftheriades et~al.}(2002)\citenamefont{Eleftheriades,
  Izer, and Kremer}}]{Eleftheriades02}
\bibinfo{author}{\bibfnamefont{G.}~\bibnamefont{Eleftheriades}},
  \bibinfo{author}{\bibfnamefont{A.}~\bibnamefont{Izer}}, \bibnamefont{and}
  \bibinfo{author}{\bibfnamefont{P.}~\bibnamefont{Kremer}},
  \bibinfo{journal}{IEEE Trans. Microwave Theory and Techniques}
  \textbf{\bibinfo{volume}{50}}, \bibinfo{pages}{2702} (\bibinfo{year}{2002}).

\bibitem[{\citenamefont{Salehi et~al.}(2005)\citenamefont{Salehi, Majedi, and
  Mansour}}]{Salehi05}
\bibinfo{author}{\bibfnamefont{H.}~\bibnamefont{Salehi}},
  \bibinfo{author}{\bibfnamefont{A.~H.} \bibnamefont{Majedi}},
  \bibnamefont{and} \bibinfo{author}{\bibfnamefont{R.~R.}
  \bibnamefont{Mansour}}, \bibinfo{journal}{IEEE Trans. Appl.
  Superconductivity} \textbf{\bibinfo{volume}{15}}, \bibinfo{pages}{996}
  (\bibinfo{year}{2005}).

\bibitem[{\citenamefont{Jung et~al.}()\citenamefont{Jung, Butz, Shitov, and
  Ustinov}}]{Jung_arXiv}
\bibinfo{author}{\bibfnamefont{P.}~\bibnamefont{Jung}},
  \bibinfo{author}{\bibfnamefont{S.}~\bibnamefont{Butz}},
  \bibinfo{author}{\bibfnamefont{S.~V.} \bibnamefont{Shitov}},
  \bibnamefont{and} \bibinfo{author}{\bibfnamefont{A.~V.}
  \bibnamefont{Ustinov}}, \bibinfo{note}{arXiv.1301.0440v1}.

\bibitem[{\citenamefont{Pozar}(2005)}]{Pozar05}
\bibinfo{author}{\bibfnamefont{D.}~\bibnamefont{Pozar}},
  \emph{\bibinfo{title}{Microwave Engineering}} (\bibinfo{publisher}{Wiley},
  \bibinfo{address}{New York}, \bibinfo{year}{2005}), \bibinfo{edition}{3rd}
  ed.

\bibitem[{\citenamefont{Ashcroft and Mermin}(1976)}]{Ashcroft76}
\bibinfo{author}{\bibfnamefont{N.}~\bibnamefont{Ashcroft}} \bibnamefont{and}
  \bibinfo{author}{\bibfnamefont{N.}~\bibnamefont{Mermin}},
  \emph{\bibinfo{title}{Solid state physics}}
  (\bibinfo{publisher}{Holt-Saunders}, \bibinfo{year}{1976}).

\bibitem[{\citenamefont{Mooij et~al.}(1999)\citenamefont{Mooij, Orlando,
  Levitov, Tian, H.~van~der Wal, and Lloyd}}]{Mooij99}
\bibinfo{author}{\bibfnamefont{J.~E.} \bibnamefont{Mooij}},
  \bibinfo{author}{\bibfnamefont{T.~P.} \bibnamefont{Orlando}},
  \bibinfo{author}{\bibfnamefont{L.}~\bibnamefont{Levitov}},
  \bibinfo{author}{\bibfnamefont{L.}~\bibnamefont{Tian}},
  \bibinfo{author}{\bibfnamefont{C.}~\bibnamefont{H.~van~der Wal}},
  \bibnamefont{and} \bibinfo{author}{\bibfnamefont{S.}~\bibnamefont{Lloyd}},
  \bibinfo{journal}{Science} \textbf{\bibinfo{volume}{285}},
  \bibinfo{pages}{1036} (\bibinfo{year}{1999}).

\bibitem[{\citenamefont{Clarke and Wilhelm}(2008)}]{Insight}
\bibinfo{author}{\bibfnamefont{J.}~\bibnamefont{Clarke}} \bibnamefont{and}
  \bibinfo{author}{\bibfnamefont{F.~K.} \bibnamefont{Wilhelm}},
  \bibinfo{journal}{Nature} \textbf{\bibinfo{volume}{453}},
  \bibinfo{pages}{1031} (\bibinfo{year}{2008}).

\bibitem[{\citenamefont{Koch et~al.}(2007)\citenamefont{Koch, Yu, Gambetta,
  Houck, Schuster, Majer, Blais, Devoret, Girvin, and Schoelkopf}}]{Koch07}
\bibinfo{author}{\bibfnamefont{J.}~\bibnamefont{Koch}},
  \bibinfo{author}{\bibfnamefont{T.~M.} \bibnamefont{Yu}},
  \bibinfo{author}{\bibfnamefont{J.}~\bibnamefont{Gambetta}},
  \bibinfo{author}{\bibfnamefont{A.~A.} \bibnamefont{Houck}},
  \bibinfo{author}{\bibfnamefont{D.~I.} \bibnamefont{Schuster}},
  \bibinfo{author}{\bibfnamefont{J.}~\bibnamefont{Majer}},
  \bibinfo{author}{\bibfnamefont{A.}~\bibnamefont{Blais}},
  \bibinfo{author}{\bibfnamefont{M.~H.} \bibnamefont{Devoret}},
  \bibinfo{author}{\bibfnamefont{S.~M.} \bibnamefont{Girvin}},
  \bibnamefont{and} \bibinfo{author}{\bibfnamefont{R.~J.}
  \bibnamefont{Schoelkopf}}, \bibinfo{journal}{Phys. Rev. A}
  \textbf{\bibinfo{volume}{76}}, \bibinfo{pages}{042319}
  (\bibinfo{year}{2007}).

\bibitem[{\citenamefont{G\"uhne and Toth}(2009)}]{Guhne09}
\bibinfo{author}{\bibfnamefont{O.}~\bibnamefont{G\"uhne}} \bibnamefont{and}
  \bibinfo{author}{\bibfnamefont{G.}~\bibnamefont{Toth}},
  \bibinfo{journal}{Physics Reports} \textbf{\bibinfo{volume}{474}},
  \bibinfo{pages}{1} (\bibinfo{year}{2009}).

\bibitem[{\citenamefont{Leggett et~al.}(1987)\citenamefont{Leggett,
  Chakravarty, Dorsey, Fisher, Garg, and Zwerger}}]{Leggett87}
\bibinfo{author}{\bibfnamefont{A.}~\bibnamefont{Leggett}},
  \bibinfo{author}{\bibfnamefont{S.}~\bibnamefont{Chakravarty}},
  \bibinfo{author}{\bibfnamefont{A.}~\bibnamefont{Dorsey}},
  \bibinfo{author}{\bibfnamefont{M.}~\bibnamefont{Fisher}},
  \bibinfo{author}{\bibfnamefont{A.}~\bibnamefont{Garg}}, \bibnamefont{and}
  \bibinfo{author}{\bibfnamefont{W.}~\bibnamefont{Zwerger}},
  \bibinfo{journal}{Rev. Mod. Phys.} \textbf{\bibinfo{volume}{59}},
  \bibinfo{pages}{1} (\bibinfo{year}{1987}).

\bibitem[{\citenamefont{Weiss}(1999)}]{Weiss99}
\bibinfo{author}{\bibfnamefont{U.}~\bibnamefont{Weiss}},
  \emph{\bibinfo{title}{Quantum Dissipative Systems}}, no.~\bibinfo{number}{10}
  in \bibinfo{series}{Series in modern condensed matter physics}
  (\bibinfo{publisher}{World Scientific}, \bibinfo{address}{Singapore},
  \bibinfo{year}{1999}), \bibinfo{edition}{2nd} ed.

\bibitem[{\citenamefont{Paauw et~al.}(2009)\citenamefont{Paauw, Fedorov,
  Harmans, and Mooij}}]{Paauw09}
\bibinfo{author}{\bibfnamefont{F.~G.} \bibnamefont{Paauw}},
  \bibinfo{author}{\bibfnamefont{A.}~\bibnamefont{Fedorov}},
  \bibinfo{author}{\bibfnamefont{C.~J. P.~M.} \bibnamefont{Harmans}},
  \bibnamefont{and} \bibinfo{author}{\bibfnamefont{J.~E.} \bibnamefont{Mooij}},
  \bibinfo{journal}{Phys. Rev. Lett.} \textbf{\bibinfo{volume}{102}},
  \bibinfo{pages}{090501} (\bibinfo{year}{2009}).

\bibitem[{\citenamefont{Cardy}(1981)}]{Cardy81}
\bibinfo{author}{\bibfnamefont{J.}~\bibnamefont{Cardy}}, \bibinfo{journal}{J.
  Phys. A: Math. Gen.} \textbf{\bibinfo{volume}{14}}, \bibinfo{pages}{1407}
  (\bibinfo{year}{1981}).

\end{thebibliography}

\end{document}